# Prostate biopsies guided by 3-dimension real-time (4D) transrectal ultrasound on a phantom. Comparative study versus 2D transrectal ultrasound guided biopsies.

JA Long, V Daanen, A Moreau-Gaudry, J Troccaz, JJ Rambeaud, JL Descotes

**Abstract**

**Objective:** The objective of the study is to evaluate the accuracy in localisation and distribution of real-time 3-dimension (4D) ultrasound guided biopsies on a prostate phantom.

**Methods**: A prostate phantom was created. A 3D real-time ultrasound system with a 5.9 MHz volumic endorectal probe was used, making it possible to see several reconstructed orthogonal viewing planes in real-time. Fourteen operators performed biopsies. Twelve cores were sampled by every operator, first under 2D then 4D TRUS guidance (336 biopsies).

The biopsy path was modelled using segmentation in a 3D ultrasonographic volume. Each biopsy path was visualised using the trail left by the needle in the phantom. Only one biopsy was performed per phantom. The 3D volumes of each sampled phantom were then manually registered, without segmentation, with a reference phantom. Special software was used to visualise the biopsy paths in the reference prostate and assess the sampled area. A comparative study was performed to determine the benefits of 4D as compared to 2D TRUS guidance, looking at the accuracy of the entry points and target of the needle. Distribution was assessed by measuring the volume sampled and a redundancy ratio of the sampled voxels.

**Results:** A significantly increase in accuracy in hitting the target zone was identified using 4D ultrasonography as compared to 2D. There was no increase in the sampled volume or improvement in the biopsy distribution with 4D ultrasonography as compared to 2D.

**Conclusion:** The 4D TRUS guidance appears to show, on a synthetic model, an improvement in location accuracy and in the ability to reproduce a protocol. The biopsy distribution does not seem improved.

# Introduction

The aim of this study is to assess the value of 3 dimensions and real-time (4D mode) ultrasonography in ultrasound-guided prostate biopsies.

We want to determine whether the ability to visualise simultaneously orthogonal planes allows a more accurate and more reproducible sampling, and whether the biopsies are more evenly distributed within the prostate volume.

Our study compares traditional 2D TRUS guided biopsy with 3D real-time (4D) TRUS guided biopsy on a synthetic model (phantom).

The objective of the study is not to assess a biopsy protocol, but rather the effectiveness of each guiding method (2D and 4D) in carrying out a specific protocol.

2D TRUS was considered a promising technique for detecting non-palpable tumor lesions. It turned out that the presence of a hypoechogenic nodule in the peripheral zone did not show sufficient sensitivity and, more importantly, specificity.

In 1989, the team from Stanford was the first to show that sextant biopsies improved a better cancer detection rate than biopsies focusing on ultrasound anomalies[1]. Despite the improvement in detection rate, it was shown that cancer was detected in 20% of the cases studied, during a second set following a negative sextant biopsy. This tended to show that the ideal biopsy scheme remained to be found and that it was difficult to determine the features of an ideal protocol [1].

This observation was modelled by Chen, who created a statistic model allowing biopsy protocol simulation, using 607 cancer specimens on computer-reconstructed radical prostatectomy specimens. This method showed that 27% of cancer specimens exceeding 0,5 cc would not have been detected using the Stanford protocol[2].

Many biopsy schemes have been described to improve the sensitivity of biopsies while limiting the number of samples. The trend is now toward an increase in the number of biopsies and sampling the peripheral zone. Additional morbidity resulting from extensive protocols remains controversial[3].

Three dimensional ultrasonography is commercially available for routine clinical use and has been used on some devices since 1994. Rather than producing an anatomic section, as is the case with traditional 2D ultrasonography, image acquisition is performed as a volume of data with nearly immediate reconstruction and simultaneous display of sectional anatomy in three orthogonal planes--sagittal plane, transverse or coronal plane, or any arbitrary oblique. As a result 3D US allows unrestricted access to an infinite number of viewing planes. The scanning time varies from 3 to 10 seconds, depending on the volume and quality desired. It is also possible to use endocavity probes, in which case the scanning can either be rotational (around the probe's axis) or angular, which is the case with the device tested (General Electric Voluson 730 pro ®).

A 3D-volume acquisition of an organ can also be performed, such that the images can be manipulated and the coordinates of all volume points stored in the memory, with the related grey levels. 3D ultrasound appears to be helpful in predicting extracapsular extension[4].

The latest developments in ultrasound have led to real-time 3D acquisition (4D-mode). It consists of displaying, simultaneously, 3 orthogonal viewing planes.

This guiding mode has proven its effectiveness in terms of accuracy in biopsies on breast tumours[5], liver tumours[6] and pancreatic pseudocysts [7]. We wish to study its benefits in prostate biopsy.

**Methods**

The device used is a VOLUSON® 730 Pro (General Electric) equipped with an RIC5-9 (5.9 MHz) endorectal probe, which can be used in 2D, 3D or 4D mode. The biopsy needle was Tru-Cut 18G, cutting length of 23 mm. It appears important to specify that the benefits of 4D ultrasound reside not in high-accuracy guiding toward a visible zone that is materialised through ultrasound, but rather in better sampling of the gland. Our purpose is not to assess the protocol, but rather the effectiveness in guiding the biopsy toward the zones theoretically set out by the protocol. Testing the device on a synthetic model appeared a necessary preliminary step.

**Creation of a synthetic prostate model (phantom)**

In a square plastic box (10 cm x 10 cm), a circular orifice spanning 6 cm in diameter was cut, so that a probe could be inserted. Inside the orifice, a cylinder intended to simulate the rectal wall was placed. The ultrasound propagation environment was made of Aquasonic® ultrasound gel (750 ml per phantom). The prostate was made by squeezing candle gel into a silicon mould. The model was the equivalent of a 40-ml prostate. Three landmarks were placed on the convex side so as to more effectively registrate the phantoms with manual methods (figure 1).

After biopsy execution, air was brought in the phantom by the needle. This made it possible to assess the biopsy path using ultrasonographic volume acquisition (3D). The same phantom model was designed for each biopsy (figure 2).

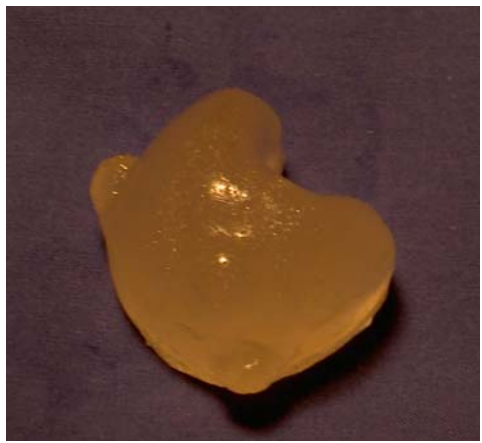

**Figure 1: Candle-gel prostate phantom**

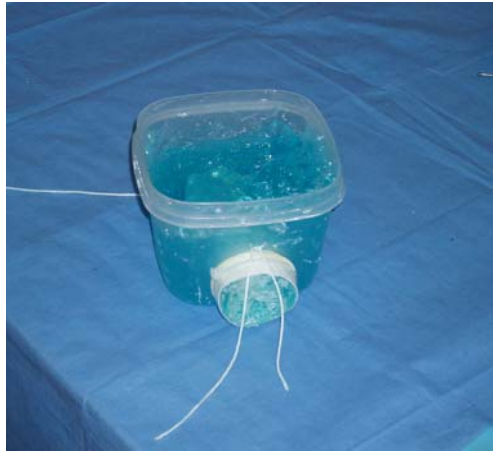

**Figure 2: Endorectal prostate biopsy phantom**

### Modelling the biopsy path

The air incorporated into the phantom when the needle entered left a hyperechogenic trail that was easily detectable by ultrasonography (figure 3). A 3D volume acquisition of the entire phantom in which the biopsy was taken was performed.

In three consecutive matching tests, it was possible to superpose the trail left by the air on that of the needle.

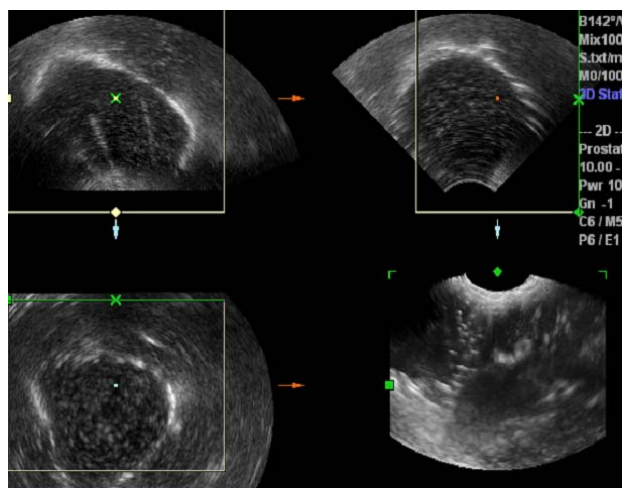

**Figure 3: Visualisation of 3 biopsy trails in the same phantom**

### Biopsy scheme

The protocol used was a 12-biopsy scheme, in which 2 biopsies (paramedian and lateral) were performed at the apex, in the middle and at the base of each prostate lobe.

Considering the bias that would have been created by the trails from the previous biopsies, it was essential that one phantom be produced per biopsy.

For that reason, 12 phantoms were produced, making it possible to perform a single uninterrupted session. The phantoms were replaced after each session and 336 phantoms were produced.

Each operator performed the first biopsy session using the VOLUSON 730® with 2D mode, as usual, then with 4D mode.

Following each biopsy, an ultrasonographic volume was acquired and recorded.

### Processing the ultrasonographic data acquired

*Producing a reference prostate.*

The first step in processing the biopsy data was to create a segmented reference volume (template), using the image processing software, Analyze®. The reference volume was produced during a 3D-acquisition using standard settings for all of the acquisitions in the trial (17.7 cm-15Hz). Following segmentation, the cloud of points was exported to a software developed specially for our manipulations on Visual C++®, making it possible to visualise the reference prostate volume.

*Registration of biopsies with the reference prostate (common reference base)*

Volume acquisition was performed on each phantom containing a single biopsy.

The volumes were matched using the manual volume registration function on Analyze®.

Attempts at automatic registration did not yield any results, due to the significant noise and significant variations in interface contrast between the various phantoms.

The biopsy trail in the phantom was manually segmented using Analyze® in the transformed image.

The twelve biopsies carried out in one session were collected and shown in the prostate reference (figures 4 and 5). The methodology is summarised in figure 6.
.

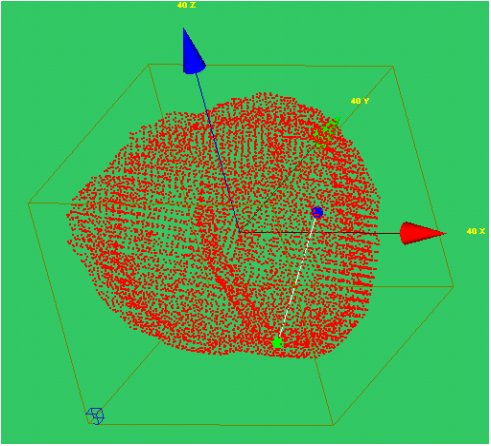

**Figure 4: Visualisation of a biopsy path in the reference volume. The green ball shows the entry point. The blue ball shows the target hit.**

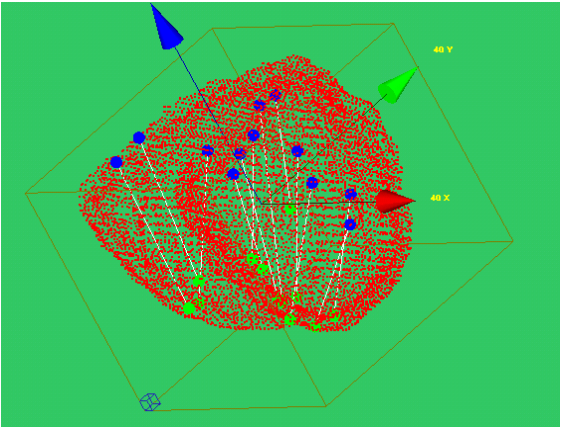

**Figure 5: Superposition of all biopsies in the template.**

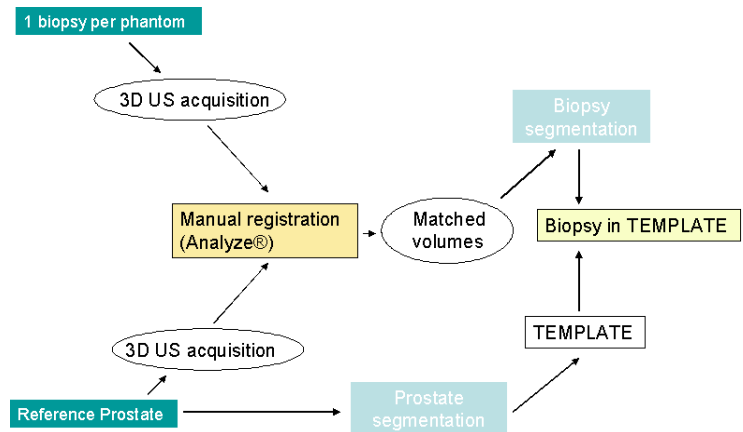

**Figure 6: Summary methodology scheme**

*Assessing biopsy results*

A quality biopsy assessment protocol was established. Three parameters were studied:

- **Entry zone**: Based on the anatomical and protocol schemes for the 12 biopsies, a point located on the prostate surface was chosen as the ideal insertion point for the needle in each biopsy. Considering that there was no clear recommendation or obvious target, an entry zone spanning 7 mm in diameter (sphere) around the said puncture point was deemed suitable. Beyond that zone the error distance was measured. The 7-mm diameter was chosen as it would allow the entry zones to be juxtaposed on the phantom surface without overlapping.

-**Target zone**: The biopsy tip was also compared to an ideal target zone of 7-mm in diameter. The error was defined as the distance separating the biopsy tip and the sphere (figure 7).

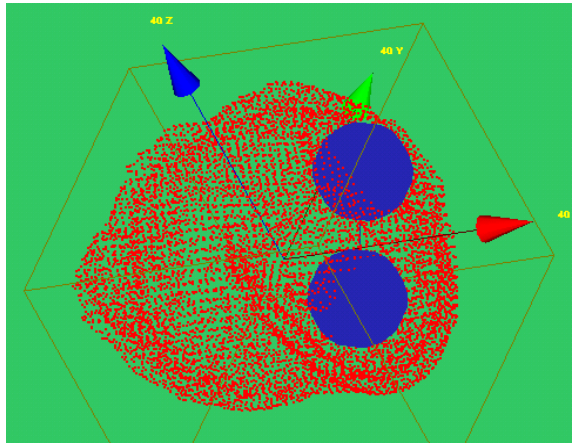

**Figure 7: Creation of ideal entry and target spheres for one biopsy.**

- **Biopsy distribution in the volume**.

Assessing the appropriate biopsy distribution within the prostate volume was not easy. According to Epstein, cancer is significant above 0.5 ml[8]. We therefore tried to find a method that would make it possible to show the volume "explored" by the biopsies (depending on biopsy length in the prostate and, above all, overlapping between the biopsies). The theoretical hypothesis was that 2 parallel biopsies 10 mm apart could not miss a significant cancer specimen (the diameter of a 0.5-cc ball is 10 mm). We therefore hypothesized that a biopsy "explored" a volume within a 1.2 mm-diameter cylinder (18 gauge needle diameter), in addition to which there was a "security zone" of 10-mm diameter.

Figure 8 shows the oversampling of a zone, in which 2 cylindrical volumes are covered by bringing together 2 biopsies. As a result, the adjacent regions of the prostate are undersampled.

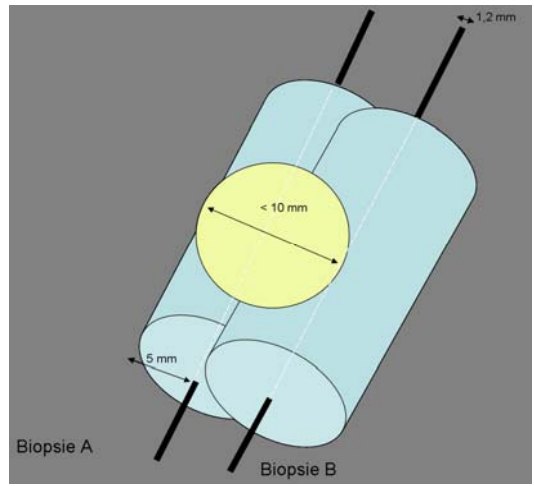

**Figure 8: Overlapping of 2 "biopsy cylinders"**

In other words, the biopsy-explored prostate volume was equal to the length of the biopsy needle multiplied by the surface of a 5.6 mm-radius disk.

Each voxel (ultrasonographic volume unit) in the biopsy cylinder was allocated a value of 1 or 0; those within the biopsy cylinder were given a value equal to one. The voxels that overlapped with the adjacent biopsy cylinders were counted only once.

We opted for a ratio: the number of voxels explored in the prostate/number of voxels that should have been explored biopsied if the biopsies did not overlap. This method indirectly reflected the distribution, by identifying the ability of each operator to allocate the biopsies so that they would not overlap.

During each session, it was possible to determine the volume identified by the series within the prostate.

*Statistical analysis*
Descriptive and inferential statistical analysis of the data was performed using STATVIEW 5.0® software, in conjunction with the Grenoble Technological Innovative Centre (TIC). The quantitative variables were described by their mean and standard deviation (Mean ± SD) with median and interquartile range. Means comparisons were performed by using suitable parametric tests once the application conditions where checked. When the said conditions were not checked, suitable non-parametric tests were used. As usual, tests results were considered statistically significant for a p-value < 0.05.

Accuracy analysis was performed on each 2D and 4D TRUS guided biopsy by analysing distances at the ideal entry zones and ideal theoretical target zones (paired t-test). The differences of distances at the entry zones (and then at the ideal target zones) between the 2D and 4D approaches were analysed taking account of location factors in the sagittal (apex, middle and base) and transversal (paramedian, lateral) sections (One factor ANalysis Of VAriance). Biopsy reproducibility analysis was performed by looking at the deviation from the mean (paired t-test). A non parametric test (Mann-Whitney test) was performed to determine the statistical influence of each biopsy guidance mode on the percentage of the prostate covered by only 1 biopsy.

The standard biopsy protocol was explained to each operator using a summary scheme. The 2D procedure was carried out first. The second session was held within 3 hours to 3 weeks.

## Results

In total, 336 biopsies were performed by 14 operators on 12 phantoms per session. The paired series tests involved 158 paired biopsies.

An increase in accuracy in hitting the target zone was identified significantly in 4D as compared to 2D (p=0.0042). In contrast, no significant improvement in biopsy accuracy was found for the entry zone (table 1 and 2).

**Table 1: Mean distances from entry and target zones in ideal protocol, in 2D and 4D modes**

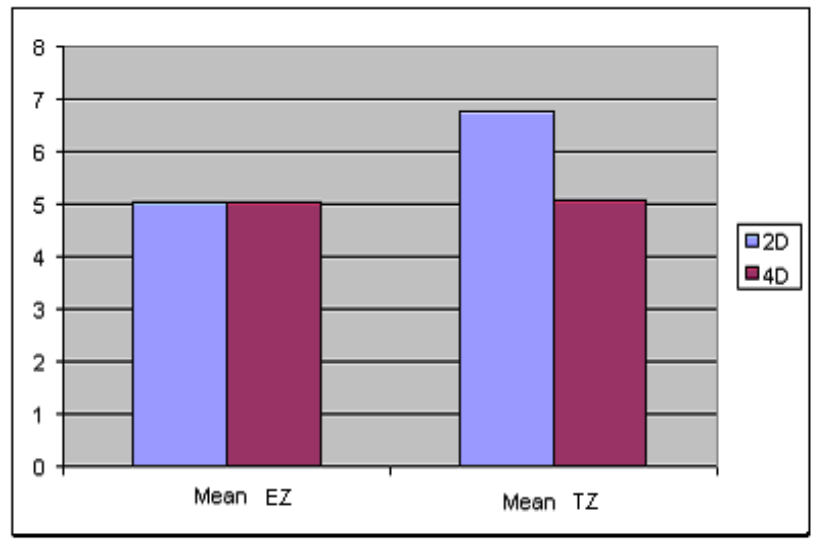

**Table 2: Summary table regarding biopsy accuracy in reaching ideal theoretical protocol**

|  | n | Procedure | Mean (mm) +/- SD | Paired Student t test |
|---|---|---|---|---|
| **Distance from ideal target zone** | 158 | **2D** | **6.79 ± 6.18** | **S** |
|  |  | **4D** | **5.1± 4.8** |  |
| **Distance from ideal entry zone** | 158 | **2D** | **5.28 ± 5.71** | NS |
|  |  | **4D** | **5.19 ± 4.83** |  |

Furthermore, the means of differences of distance between 2D and 4D approach, in accuracy in hitting target zone (and then at the entry zone), were not statistically different at each prostate location, as shown by one factor analysis (table 3 and 4).

**Table 3: Interaction curve between differences in accuracy in hitting target zone in 2D and 4D, by location on prostate.**

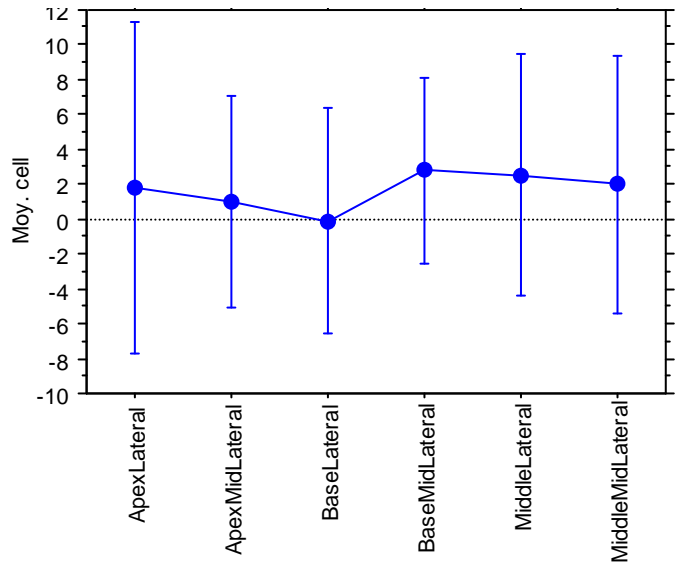

**Table 4: Interaction curve between differences in accuracy at the entry zone in 2D and 4D, by prostate zone.**

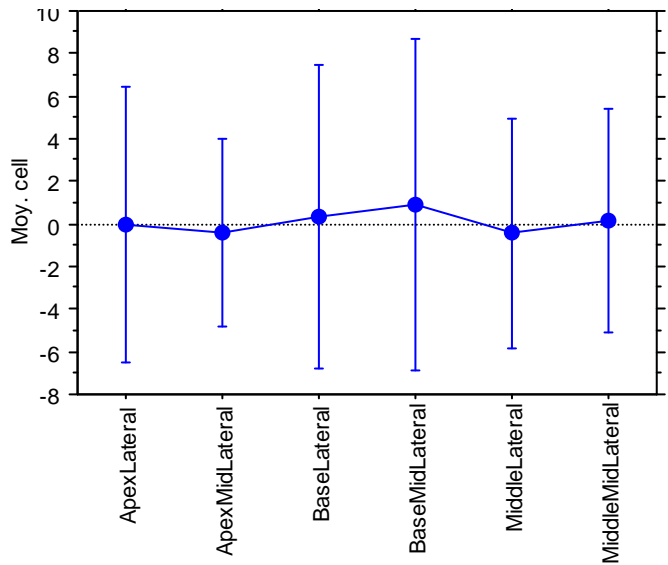

A biopsy reproducibility test was performed by looking at the variance in distances at the entry and target points. It showed that reproducibility increased significantly in biopsies toward the target zone with 4D US mode (p= 0.0385). However, there is no statistical link, but a trend toward improvement in reproducibility in needle entry points in the phantom (p=0.0623) (table 5).

**Table 5: Biopsy reproducibility study by deviation from mean² (variance)**

|  | n | Procedure | Deviation from mean (mm) | Standard deviation | Student Test |
|---|---|---|---|---|---|
| **Variance in distance compared to ideal target zone** | 158 | **2D** | **37.95** | 78.45 | **0.0385 S** |
|  |  | **4D** | **22.9** | 41.38 |  |
| **Variance in distance compared to ideal entry zone** | 158 | **2D** | **32.43** | 59.66 | 0.0623 NS |
|  |  | **4D** | **23.18** | 33.15 |  |

The biopsy distribution data did not show any significant difference in the prostate volume explored under 2D and 4D US guidance (p=0.44), with a mean of 1025 mm³ per biopsy (12,3cc) per 4D guided biopsy session), compared to 971 mm³ under 2DUS guidance (11,65 cc per 2D guided biopsy session). The sampled volume percentage not covered in several biopsies (sampled voxels through which only one biopsy runs) was 56.9% under 2D guidance, as compared to 56.7% under 4D guidance (p=0.98) (table 6).

**Table 6: Study of biopsy dispersion by measuring volume per biopsy and percentage of the prostate sampled by only one biopsy**

|  | N | Procedure | (median, (interquartile range) | mean ± SD | p |
|---|---|---|---|---|---|
| **Volume sampled per biopsy** | 14 | 2D | 963 (209) | 971 mm³ ±166 | NS |
|  |  | 4D | 907 (258) | 1025 mm³±184 |  |
| **Percentage of sampled prostate** | 14 | 2D |  | 0.569 |  |
| **through which only one biopsy runs** | | 4D | | 0.567 | NS |

## Discussion

We have observed that, on a phantom model, biopsy accuracy on the zones to be sampled improved with 4D guidance. This illustrates the fact that multi-dimensional visualisation improves the operator's ability to hit the virtual target. The improved reproducibility in execution shows the robustness of the guidance. In contrast, no benefits were found when 3D real time reconstruction was used as regards a possible increase in volume sampled by the set and in biopsy dispersion. This would have been a powerful claim for improving the sensitivity of a series of biopsies. Subjectively, it would appear that the 4D guiding technique is more intuitive and more beneficial for less-experienced operators, but no tie has been demonstrated. A number of biases were found throughout the work methodology. The limits inherent in the phantom material do not exactly replicate the tactile conditions of performing a biopsy. More importantly, the registration (matching volumes) introduced a bias in assessing the position of the points segmented in the biopsy. We saw that only biopsy was performed per phantom, to prevent the operator from being able to see his previous biopsies and benefiting from their guidance. For that reason, at each session, 12 registrations had to be performed. As a result, the risk of bad matching was 12 times greater. Lastly, manual registration does not allow for a perfectly reproducible method. However, those biases should cancel each other out in that they are present whatever the guidance method.

The various automatic registrations techniques developed have not, for the time being, yielded significant results. A registration algorithm is still being developed.

It is not easy to find high-quality criteria for a series of prostate biopsies: the purpose of the biopsies is to detect the tumours. The current TRUS guided biopsy technique involves randomising biopsies inside the prostate. In most cases, there is no visible target. We set out two objectives: to be as accurate as possible in producing a selected biopsy protocol and to scatter the biopsies so that they would not overlap. The ability to produce the most accurate protocol possible does not, then, assess the ability of the biopsies to detect cancer, but rather the ability to perform a selected protocol.

A more judicious criterion would be to assess the allocation of samples in the prostate volume, meaning to determine the ability to spread the biopsies across the prostate, without performing a biopsy on an already-sampled area. We assumed that a biopsy "sampled" a volume equal to the inside of a 5.6 mm-radius cylinder. This volume reflects clinical reality – that of a cancer specimen of significant volume, as currently defined by Epstein (volume of 0.5 cc, Gleason score > 6). Our distribution assessment technique explores coverage, but does not actually look at how the biopsies spread across the volume (figure 9). Both A and B biopsies achieved a maximum non-coverage rate, but ignore a cancer specimen of significant size.

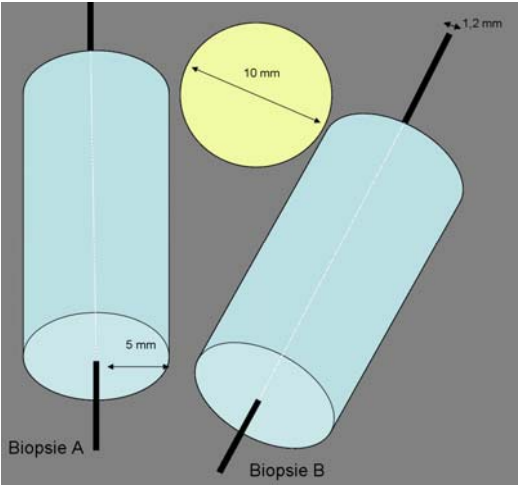

**Figure 9: Faulting our distribution assessment method. A significant-sized cancer goes undetected between 2 biopsy cylinders that do not overlap in the prostate volume.**

Consequently, a biopsy session should be judged first on accuracy in placing the needle where planned and secondly the ability to not have the needles enter through the same points. This studied showed that the first point was improved under 4D guidance, but not the second. Regarding accuracy, the results significantly prove that concurrent visualisation of several viewing planes improves the spatial representation of prostate biopsy. In contrast, the low number of operators (14) makes interpreting the results a thorny task as regards biopsy distribution analysis. No significant difference was shown between 2D and 4D guidance.

**The Prospects for 3D Ultrasound in Diagnosing Prostate Cancer**

- **Memorising biopsy paths**

The same approach to visualising biopsy paths could also apply to living organisms in order to establish a tumour map inside the prostate. The difference in living organisms lies mainly in how to assess the needle path. The trail left by the needle from the haematoma is a haphazard indication. It is therefore mandatory that volume acquisition be performed with the needle in place. Registrating the various biopsy needles would make it possible to display in a common reference base the path of all biopsies that might be segmented manually or more automatically.

The clinical benefit of establishing a biopsy mapping is clear in 2 situations: in a histologically suspicious zone around which additional samples may be taken, or repeated biopsies in response to a high or growing PSA when the first series was negative. In those cases, there would be no use in targeting previously biopsied zones, but rather to turn to other areas.

- **Real-time guiding of prostate biopsies**:

TIMC Laboratory is developing a biopsy guiding system to a pre-procedure planning. The planning could focus on an optimised protocol[9] or on areas previously sampled by a first set of biopsies. A suspicious zone requiring a repeat biopsy could become the target for a guidance system or a special schedule could be applied (biopsies of the transition zone or biopsies of non-sampled zones during a first set, or interspersing the biopsy path between previous biopsies). The aim is to detect cancer specimens that are smaller or less detectable during standard biopsy procedures. The targeted objective schedule differs from recently-proposed saturation biopsy protocols that involve performing a larger number of biopsies in the gland in an attempt to cover as much volume as possible.

The final step is to be able to carry out a pre-procedure planning focusing on assessment imaging techniques (spectroscopy and injected IRM) and to perform real-time ultrasound guidance after data registration. For the time being, it appears difficult to register imaging procedures without segmentation and real-time matching of the 2 techniques might be a problem, due in large part to processor computation times. The first attempts at computer-assisted guidance, developed at TIMC Laboratory, is currently being tested at the Grenoble University Hospital Centre [10].

## Conclusion:

The study is a step in pre-operative assessment of prostate biopsy paths using 3D ultrasound. It also makes it possible to demonstrate that real-time 3D guidance on phantom biopsies offers greater accuracy in determining the needle target and makes biopsy reproducibility better. No evidence has been shown that biopsy distribution is more even in the prostate volume or that the biopsies were better spread.

**Acknowledgments:**
Our thanks to the French Association of Urology (AFU), Sanofi-Aventis for the grant awarded, and the nationwide PHRC Prostate Echo 2003 project, which made this work possible.